\documentclass[12pt]{article}

\usepackage{graphics, epsf%, psfig
}

\newcommand{\nit}{\noindent}

\newcommand{\np}{\newpage}
\newcommand{\dsp}{\displaystyle}
\newcommand{\vs}[1]{\vspace{#1 ex}}
\newcommand{\hs}[1]{\hspace{#1 em}}
\newcommand{\bfr}{\begin{flushright}}
\newcommand{\efr}{\end{flushright}}
\newcommand{\bc}{\begin{center}}
\newcommand{\ec}{\end{center}}
\newcommand{\ben}{\begin{enumerate}}
\newcommand{\een}{\end{enumerate}}

\newcommand{\be}{\begin{equation}}
\newcommand{\ee}{\end{equation}}
\newcommand{\bea}{\begin{eqnarray}}
\newcommand{\eea}{\end{eqnarray}}
\newcommand{\ba}{\begin{array}}
\newcommand{\ea}{\end{array}}
\newcommand{\ct}{\cite}
\newcommand{\bit}{\bibitem}
\newcommand{\dd}[2]{\frac{\partial{#1}}{\partial{#2}}}

\newcommand{\bg}{\beta}
\newcommand{\gam}{\gamma}
\newcommand{\del}{\delta}
\newcommand{\eps}{\epsilon}

\newcommand{\zg}{\zeta}
\newcommand{\thg}{\theta}

\newcommand{\Fg}{\Phi}
\newcommand{\Sg}{\Sigma}
\newcommand{\Og}{\Omega}
\newcommand{\Lb}{\Lambda}

\newcommand{\bfa}{{\bf a}}

\newcommand{\bfA}{{\bf A}}

\newcommand{\bfP}{{\bf P}}

\newcommand{\lh}{\left(}
\newcommand{\rh}{\right)}

\newcommand{\cH}{{\cal H}}

\begin{document}

\pagestyle{empty}
\begin{flushright}
NIKHEF/2005-005
\end{flushright}
\vs{2}

\begin{center}
{\Large{\bf{A note on BRST quantization}}} \\
\vs{2}
{\Large {\bf{of $SU(2)$ Yang-Mills mechanics}}} \\
\vs{7}

{\large A.\ Fuster$^{*}$ and J.W.\ van Holten$^{\dagger}$} \\
\vs{2}

{\large{NIKHEF, Amsterdam NL}} \\
\vs{7}

\today
\end{center}
\vs{10}

\nit
{\small
{\bf Abstract} \\
The quantization of $SU(2)$ Yang-Mills theory reduced to $0+1$ space-time
dimensions is performed in the BRST framework. We show that in the unitary
gauge $A_0 = 0$  the BRST procedure has difficulties which can be solved
by introduction of additional singlet ghost variables. In the Lorenz gauge
$\dot{A}_0 = 0$ one has additional unphysical degrees of freedom, but the
BRST quantization is free of the problems in the unitary gauge. }
\vfill

\footnoterule
\nit
{\footnotesize{
$^*$ e-mail: {\tt fuster@nikhef.nl} \\
$~~^{\dagger}$ e-mail: {\tt v.holten@nikhef.nl} }}

\np
~\hfill

\np
\pagestyle{plain}
\pagenumbering{arabic}

\section{$SU(2)$ Yang-Mills mechanics \label{s.1}}

We consider $SU(2)$ Yang-Mills mechanics obtained by the reduction
of $SU(2)$ Yang-Mills field theory in $(D+1)$-dimensional space-time
to a finite-dimen\-sional quantum system, by taking the dynamical variables
to depend on the time co-ordinate $t$ only. The lagrangean of such a
theory is
\be
L_{\mbox{\tiny YMQM}} = \frac{1}{2}(F_{0i}^a)^2 - \frac{1}{4}\, (F_{ij}^{a})^2,
\label{p.1}
\ee
where $i, j = (1,...,D)$, and
\be
F_{0i}^a = \dot{A}_i^a - g \eps^{abc} A_0^b A_i^c , \hs{2}
F_{ij}^a = - g \eps^{abc} A_i^b A_j^c.
\label{p.2}
\ee
Such a system has been widely studied in the context of non-perturbative
aspects of (super) Yang-Mills theories \ct{luscher,claudson-halpern}, and 
as a first step in the regularized dynamics of membrane theory 
\ct{halpern-schwartz}-\ct{fghhy}.

The lagrangean is invariant under time-dependent gauge transformations
with parameters $\Lb^a(t)$, taking the infinitesimal form
\be
\del A_0^a = \dot{\Lb}^a - g \eps^{abc} A^b_0 \Lb^c, \hs{2}
\del A_i^a = - g \eps^{abc} A_i^b \Lb^c.
\label{p.3}
\ee
This invariance allows us to impose a gauge condition leaving the physical
dynamics unchanged. The simplest choice is
\be
A_0^a = 0.
\label{p.4}
\ee
With this condition the effective lagrangean for the remaining $D$-dimensional
vector potentials $\bfA_a$ becomes\footnote{We do not distinguish between upper
and lower adjoint indices $(a,b,c,...)$ for $SU(2)$.}
\be
L_{eff} = \frac{1}{2}\, \dot{{\bfA}}_a^2 - V[\bfA],
\label{p.5}
\ee
with the potential
\be
V[\bfA] = \frac{g^2}{4}\, \lh \bfA_a^2\, \bfA_b^2  - (\bfA_a \cdot \bfA_b)^2 \rh.
\label{p.6}
\ee
In addition, we have to impose a set of (first-class) constraints corresponding
to the previous equations of motion for $A_0^a$:
\be
G^a \equiv g\, \eps^{abc}\, A^b_i F_{0i}^c \simeq g\, \eps_{abc}\, \bfA_b \cdot
\dot{\bfA}_c = 0
\label{p.7}
\ee
Thus, the physical trajectories in configuration space in the gauge (\ref{p.4})
are the solutions of the Euler-Lagrange equations derived from (\ref{p.5})
subject to the additional constraints (\ref{p.7}).

In addition to the pure Yang-Mills theory described by the action (\ref{p.1}),
one can also construct various supersymmetric extensions, based on the
reduction of supersymmetric Yang-Mills field theory in $D = 1, 3, 5, 9$.
The spectra of these theories are qualitatively different
\ct{luscher,dewit-luscher-nicolai,halpern-schwartz,wosiek,wosiek-camp,vanbaal},
but for the problem addressed in this note those differences are not relevant.
\vs{1}

\nit
To keep track of the constraints, especially in the context of the Yang-Mills
quantum theory, we follow the BRST procedure\footnote{For reviews, 
see \ct{jwvh1} and \ct{hen-teit}.}. Thus we introduce anti-commuting ghost
degrees of freedom $(b^a, c^a)$ as well as commuting auxiliary scalars
$N^a$ in such a way, that the total gauge-fixed action becomes invariant
under a set of special ghost-dependent gauge transformations, the rigid
BRST invariance. The anti-commuting BRST differentials $\del_{\Og}$ are
defined before gauge fixing as follows
\be
\ba{ll}
\del_{\Og} A_{0}^a = (D_0 c)^a = \dot{c}^a - g \eps^{abc}\, A_0^b c^c, &
\del_{\Og} A_i^a = (D_i c)^a = - g \eps^{abc} A_i^b c^c, \\
 \\
%\del_{\Og} \lb^a = g \eps^{abc} c^b \lb^c, &
%\del_{\Og} \blb^a = -\,\overline{(\del_{\Og} \lb^a)} = g \eps^{abc} c^b \blb^c, \\
% \\
\del_{\Og} c^a = \frac{g}{2}\,\eps^{abc} c^b c^c, & \\
 \\
\del_{\Og} b^a = i N^a, & \del_{\Og} N^a = 0.
\ea
\label{p.8}
\ee
The gauge-invariance of the classical action (\ref{p.1}) implies its invariance
under the BRST transformations by construction. The BRST differential has
the standard property that $\del^2_{\Og} = 0$.
The implementation of the BRST construction for the gauge $A_0 = 0$
is, to impose this gauge condition using the Nakanishi-Lautrup fields
$N^a$ as Lagrange multipliers, and complete the effective lagrangean
so as to make it fully BRST invariant. For the case at hand this results
in the effective lagrangean
\be
L_{eff} = L_{YMQM} + N^a A_0^a
 + i b^a \lh \dot{c}^a - g \eps^{abc}\, A_0^b c^c \rh.
\label{p.10}
\ee
We can use the gauge condition implied by the Nakanishi-Lautrup fields to
eliminate $A^a_0$ and $N^a$ simultaneously; in a path-integral formulation,
this implies integrating out a $\del$-functional $\del(A_0)$. The result is
\be
L_{eff} = \frac{1}{2}\, \dot{\bfA}_a^2 - \frac{1}{4}\, (F_{ij}^a)^2
 + i b_a \dot{c}_a,
\label{p.11}
\ee
Note that for $D = 3$ we can construct a magnetic field by $\frac{1}{2} \eps_{ijk}
F_{jk}^a = B^a_i$, but this does not hold for a general $D$. The effective
lagrangean (\ref{p.11}) is invariant under the reduced form of the BRST
variations (\ref{p.8}) obtained by taking $A_0^a = 0$, and using
the equation of motion for $N^a$:
\be
\del b^a = i N^a \simeq i g \eps^{abc} \lh A_i^b F_{0i}^c - i c^b b^c \rh.
\label{p.12}
\ee
The BRST invariance of the effective lagrangean implies an anti-commuting
conserved charge by Noether's theorem. The BRST charge takes the form
\be
\Og = c^a G^a - \frac{ig}{2}\, \eps^{abc} c^a c^b b^c.
\label{p.14}
\ee
The first-class constraints of the classical theory are summarized effectively
by the statement that $\Og = 0$;  more precisely, in the phase-space formulation,
all brackets of physical quantities with $\Og$ must vanish: physical quantities
must be BRST invariant; this is discussed in more detail in the next section.

\section{Quantum theory \label{s.2}}

In the quantum theory the dynamical variables $A_i^a$ and their conjugate
momenta $P_i^a = \dot{A}_i^a$, as well as the Faddeev-Popov ghosts
are operators satisfying (anti-)commutation relations
\be
\left[ A_i^a, P_j^b \right] = i\, \del^{ab}\, \del_{ij}, \hs{2}
\left[ c^a, b^b \right]_+ = \del^{ab}
\label{p2.1}
\ee
The hamiltonian is given by
\be
H_{eff} = \frac{1}{2}\, \bfP_a^2 + \frac{1}{4}\, F_{ij}^{a\,2},
\label{p2.2}
\ee
as for pure Yang-Mills theory, in fact. The hamiltonian determines the time-evolution of any quantity $X$ constructed from the
Yang-Mills or ghost operators by the Schr\"{o}dinger equation
\be
 \dot{X} = i \left[ H, X \right].
\label{p2.3}
\ee
Gauge transformations on $(\bfA_a, \bfP_a)$ are generated by the
$SU(2)$ charges
\be
G_a = g\, \eps_{abc}\, \bfA_b \cdot \bfP_c,
\label{p2.4}
\ee
such that
\be
\del_a X =  i \left[ G_a, X \right], \hs{2}
\del_a G_b = i \left[ G_a, G_b \right] = - g\, \eps_{abc}\, G_c,
\label{p2.5}
\ee
whilst more generally the BRST transformations are given by
\be
\del_{\Og} X = i \left[\Og, X \right]_{\pm},
\label{p2.6}
\ee
the sign depending on the fermionic parity of the quantity $X$:
+ (anti-commutator) for fermionic $X$, and $-$ (commutator) for
bosonic $X$. In particular, the commutation relation (\ref{p2.5}) for
the gauge charges together with the ghost anti-commutator (\ref{p2.1})
implies the nilpotency of the BRST charge:
\be
\Og^2 = 0.
\label{p2.7}
\ee
To complete the theory we have to define an inner product on the
extended state space, such that zero-norm states decouple and physical
states have positive norm. For this to happen, it is necessary that
the BRST operator is self-adjoint w.r.t.\ this inner product. In the
co-ordinate representation, with states being represented by
wave functions $\Psi[\bfA, c]$, such an inner product is
defined by the integral \ct{jwvh2}
\be
\lh \Fg, \Psi \rh = i \int dc^1 dc^2 dc^3\, \int \prod_{i,a} dA_i^a\,
 \Fg^{\dagger}[\bfA,c] \Psi[\bfA,c].
\label{p2.8}
\ee
It is easily seen that with this definition the ghost operators
$(b^a,c^a)$ are self-adjoint themselves. It follows directly that, indeed,
\be
\lh \Og\, \Fg, \Psi \rh = \lh \Fg, \Og\, \Psi \rh.
\label{p2.8.1}
\ee

\section{Physical states \label{s.3}}

The physical states of Yang-Mills quantum mechanics are constructed
by solving for the eigenstates and eigenvalues of the hamiltonian (\ref{p2.2})
subject to the constraint of BRST invariance. \\
\nit
One useful way to construct states is by the Fock-space approach
\ct{wosiek}, in which one starts with an oscillator basis for the dynamical
degrees of freedom defined by
\be
\bfa_a^{\,} = \frac{1}{\sqrt{2}} \lh \bfA_a + i \bfP_a \rh, \hs{2}
\bfa_a^{\dagger} = \frac{1}{\sqrt{2}} \lh \bfA_a - i \bfP_a \rh.
\label{p2.9}
\ee
These creation and annihilation operators satisfy the standard commutation
relations
\be
\left[ \bfa^{\,}_a, \bfa^{\dagger}_b \right] =\del_{ab}\, 1_D,
\label{p2.10}
\ee
where $1_D$ is the $D$-dimensional unit matrix. As implied by eq.\
(\ref{p2.1}) the ghost operators already behave like fermionic ladder
operators. One is free to consider either $c^a$ or $b^a$ as creation operator; we choose $c^a$. Fock states are now constructed as polynomials
in $\bfa^{\dagger}_a$ and $c^a$ acting on an empty state $\Psi_0$
defined by
\be
\bfa^{\,}_a \Psi_0 = b^a\, \Psi_0 = 0.
\label{p2.11}
\ee 
Such a construction differs from the standard (bosonic or fermionic) creation and annihilation operators in that $c^a$ and $b^a$ are self-adjoint w.r.t. the inner product (\ref{p2.8}) rather than adjoint to each other. A similar treatment of ghost ladder operators can be found in~\cite{Grensing:2001dc}. \\

\nit
The hamiltonian can be represented as a matrix in a basis of Fock states. Subsequent diagonalization would give the spectrum of the theory\footnote{Note one can only construct a basis of finite dimension and therefore any results would be approximate (see~\cite{wosiek}).}. \\

\nit
In the context of the co-ordinate representation this construction is
realized by taking
\be
\bfa^{\,}_a = \frac{1}{\sqrt{2}} \lh \bfA_a + \dd{}{\bfA_a} \rh, \hs{2}
 b^a = \dd{}{c^a},
\label{p2.12}
\ee
and
\be
\Psi_0 = N e^{- \frac{1}{2}\, \bfA_a \cdot \bfA_a},
\label{p2.13}
\ee
with $N$ a normalization factor. In this representation  the gauge
generators are of the form
\be
G_a = - i g\, \eps_{abc}\, \bfa^{\dagger}_b \cdot \bfa_c.
\label{p3.3.1}
\ee
We analyze next the restrictions imposed by the BRST symmetry
on states in order to be physical. In the BRST formalism physical 
states are identified with the cohomology classes of the nilpotent 
BRST charge $\Og$:
\be
\cH^{phys} \simeq \frac{\mbox{Ker}\, \Og}{\mbox{Im}\, \Og}.
\label{p3.4}
\ee
This implies that physical states $\Psi$ are BRST-invariant:
\be
\Og \Psi = 0, \hs{2} \lh \Psi, \Psi \rh = 1,
\label{p3.5}
\ee
and state vectors differing by a BRST-exact state are  
identified:
\be
\Psi \sim \Psi^{\prime} = \Psi + \Og \Lb.
\label{p3.7}
\ee 
Therefore matrix elements of physical operators between physical states 
must be invariant under the BRST transformations (\ref{p3.7}):
\be
\lh \Fg, X \Psi \rh = \lh \Fg, X \Psi^{\prime} \rh, \hs{1} \mbox{if} \hs{1}
 \left[ \Og, X \right]_{\pm} = 0.
\label{p3.8}
\ee
These properties are guaranteed if BRST-exact states of the form $\Og \Lb$ 
decouple from the physical state space and have zero norm:
\be 
\lh \Psi, \Og\, \Lb \rh = \lh\Og\, \Psi, \Lb \rh = 0, \hs{2} 
\lh \Og\, \Lb, \Og\, \Lb \rh = \lh \Lb, \Og^2 \Lb \rh = 0.
\label{p3.9}
\ee
Observe that it is crucial for these results that the BRST charge is
self-adjoint w.r.t.\ to the physical inner product. \\

\nit
To do any practical calculation one needs an explicit expression
for the physical state vectors; this can be achieved by selecting
one element from each equivalence class, using the nilpotent co-BRST
operator
\be
^*\Og = G^a b^a - \frac{ig}{2}\, \eps^{abc}\, c^a b^b b^c, \hs{2}
 ^*\Og^2 = 0.
\label{p3.10}
\ee
Indeed, the co-BRST condition
\be
^*\Og \Psi = 0,
\label{p3.11}
\ee
acts as a gauge fixing condition for the BRST transformations (\ref{p3.7}),
reducing the state space as required \ct{jwvh2}. States satisfying both 
$\Omega \Psi={^*\Omega} \Psi=0$ are called BRST harmonic. Physical states 
are defined as BRST harmonic states of finite norm. We build first Fock 
states which are BRST harmonic. \\

\nit
Define the (total) ghost number as the operator 
\be
N_g=c^ab^a
\ee
Splitting the Fock space in four sectors corresponding to the 
eigenvalues $n_g$ of $N_g$: $0,\ldots,3$, we construct states 
in each ghost sector as follows,
\bea
\Psi^{(0)}[M] &=&  
 M[\bfa^{\dagger}]\, \Psi_0, \nonumber \\
\Psi^{(1)}_a[M]& =&  c_a \,
 M[\bfa^{\dagger}]\, \Psi_0, \nonumber \\
\tilde{\Psi}^{(2)}_a[M] &=&  \frac{1}{2}\,\epsilon_{abc}\, c^b c^c \,
 M[\bfa^{\dagger}]\, \Psi_0,  \nonumber \\
\tilde{\Psi}^{(3)}[M] &=& \frac{i}{3!} \;\epsilon_{abc}\; c^a c^b c^c\,
 M[\bfa^{\dagger}]\, \Psi_0, \nonumber
\eea  
Here $M[\bfa^{\dagger}]$ is some gauge-invariant polynomial in the
operators $\bfa^{\dagger}$:
\be
M[\bfa^{\dagger}] = \sum_n\, \mu_{a_1 ... a_n}\, \bfa^{\dagger}_{a_1} ...
 \bfa^{\dagger}_{a_n},
\label{p3.13}
\ee
and the coefficients $\mu_{a_1 ... a_n}$ are invariant $SU(2)$ tensors. \\

\nit
The complete set of solutions consists of two distinct classes: the 
states at ghost number $n_g = 0$, $\Psi^{(0)}[M]$, and those at ghost 
number $n_g = 3$, $\tilde{\Psi}^{(3)}[M]$. We discuss next the possibility for 
these states to have finite norm (see also~\cite{marnelius2, 
Batalin:1994rd}). The spectrum of the hamiltonian in such a basis 
would be guaranteed to be physical.

\section{Inner product and ghost vacuum \label{s.4}}

The existence of two classes of BRST-harmonic states at different
ghost number is of crucial importance for the construction of a
non-trivial physical inner product \ct{hen-teit,marnelius2}. Indeed, 
if we would only have the states at $n_g = 0$ it is quite obvious 
from the definition (\ref{p2.8}) that the vacuum state $\Psi_0$
would have zero norm:
\be
\lh \Psi_0, \Psi_0 \rh = 0,
\label{p3.14}
\ee
whilst the BRST-invariant 3-ghost operator has a non-zero vacuum
expectation value:
\be
\frac{i}{3!}\, \eps^{abc} \langle  c^a c^b c^c \rangle
 = \frac{i}{3!} \lh \Psi_0, \eps^{abc}\, c^a c^b c^c \Psi_0 \rh = 1.
\label{p3.15}
\ee
The problem clearly is in the definition of the ghost vacuum, in
combination with the fact that the ghosts are self-conjugate. Therefore
the ghost creation operators $c^a$ do not act as annihilation operators
on the conjugate (bra) vectors; if they would, the BRST charge
wouldn't be self-adjoint. In particular, it is not an option to 
replace the space of bra states by the BRST-dual states
\be
\tilde{\Psi}^{\dagger}[M] =
 \frac{i}{3!}\,  \Psi^{\dagger}[M]\, \eps^{abc}\, c^a c^b c^c,
\label{p3.16}
\ee
as proposed in \ct{marnelius2, marnelius1}, which is equivalent to the
replacement of the inner product (\ref{p2.8}) by
\be
\lh \Fg, \Psi \rh \hs{1} \rightarrow \hs{1}
 \langle \Fg, \Psi \rangle = \frac{i}{3!}\, \eps^{abc} \lh \Fg, c^a c^b c^c\, \Psi \rh.
\label{p3.17}
\ee
In fact, it is clear that the ghost variables have vanishing matrix elements
between {\em any} states (physical or unphysical):
\be
\langle \Fg, c^a \Psi \rangle = 0, \hs{2} \forall\, \Psi, \Fg,
\label{p3.18}
\ee
i.e.\ the ghosts would effectively vanish as operators, and the same is true
for the BRST charge $\Og$.

Part of the solution of this problem, also along lines suggested in 
\ct{marnelius1}, is to use the existence of the second set of solutions 
of the BRST- and co-BRST constraints with $n_g = 3$ to change the 
definition of the ghost vacuum. If we define a new vacuum state
\be
\Psi_+ = \frac{1}{\sqrt{2}} \lh 1 + \frac{i}{3!}\, \eps^{abc} c^a c^b c^c \rh \Psi_0,
\label{p3.19}
\ee
with corresponding physical excited states $\Psi_+[M] = M[\bfa^{\dagger}]\,
\Psi_+$, the ghost operators remain self-adjoint and the vacuum is
normalizable:
\be
\lh \Psi_+, \Psi_+ \rh = 1.
\label{p3.20}
\ee 
A draw-back is, that the vacuum $\Psi_+$ has no well-defined ghost number,
and not even a well defined Grassmann parity, being a sum of an even and
odd ghost number state. Moreover, the vacuum expectation value of the
ghosts (\ref{p3.15}) is changed, but still non-vanishing; actually we now have
\be
\frac{i}{3!}\, \eps^{abc}\, \langle  c^a c^b c^c \rangle_+
 = \frac{i}{3!} \lh \Psi_+, \eps^{abc}\, c^a c^b c^c \Psi_+ \rh = \frac{1}{2},
\label{p3.21}
\ee
and similarly
\be
 \frac{i}{3!}\, \eps^{abc}\, \langle b^a b^b b^c \rangle_+  = \frac{1}{2}.
\label{p3.21.0}
\ee
Although these expectation values are BRST-invariant, they carry a
non-zero ghost number, a manifestation of the non-invariance of both
the vacuum and the inner product itself under ghost rescaling.

Both problems can be solved by introducing a fourth ghost $\thg$, with
conjugate anti-ghost $\zeta$:
\be
\left[ \thg, \zeta \right]_+ = 1.
\label{p3.21.1}
\ee
The new ghost $\thg$ is taken to be a BRST singlet and has ghost number
$n_g(\thg) = - 3$; thus it has the same quantum numbers as the invariant
anti-ghost operator, whilst $\zeta$ has the quantum numbers of the
corresponding ghost operator:
\be
\thg\, \sim\, \frac{i}{3!}\, \eps^{abc}\, b^a b^b b^c, \hs{2}
\zeta\, \sim\, \frac{i}{3!}\, \eps^{abc} c^a c^b c^c.
\label{p3.22}
\ee
We then define the physical vacuum state
\be
\Fg_0 = \frac{1}{\sqrt{2}} \lh 1 + \frac{1}{3!}\, \thg\,
 \eps^{abc}\, c^a c^b c^c \rh \Psi_0,
\label{p3.23}
\ee
and the physical  excited states
\be
\Fg[M] = M[\bfa^{\dagger}]\, \Fg_0.
\label{p3.23.1}
\ee
These physical states have a well-defined ghost number $n_g(\Fg[M]) = 0$
and Grassmann parity (even). This is especially important in the 
supersymmetric extensions of the theory, as the action of the gaugino 
operators would otherwise cause problems with sign-changes for odd ghost 
number terms.

Simultaneously we also redefine the inner product (\ref{p2.8}) in the 
co-ordinate representation on the full state space to
\be
\lh \Fg, \Psi \rh = \int d\thg\, \int dc^1 dc^2 dc^3\, \int \prod_{i,a} dA_i^a\,
\Fg^{\dagger}[\bfA, c]\, \Psi[\bfA, c].
\label{p3.24}
\ee
W.r.t.\ this inner product all ghosts, including the new singlet ghost, are
self-adjoint, and so is the BRST charge $\Og$. Observe, that the ghost
integration measure now has vanishing ghost number as well.
Finally, the 3-ghost operator vacuum expectation value vanishes trivially:
\be
\frac{i}{3!}\, \eps^{abc} \langle  c^a c^b c^c \rangle_{\thg}
 = \frac{i}{3!} \lh \Fg_0, \eps^{abc}\, c^a c^b c^c \Fg_0 \rh = 0.
\label{p3.25}
\ee
Of course, there arise new vacuum expectation values
\be
\frac{1}{3!} \langle \thg\, \eps^{abc}\, c^a c^b c^c \rangle_{\thg} =
\frac{1}{3!} \langle \zeta\, \eps^{abc}\, b^a b^b b^c \rangle_{\thg} =
 \frac{1}{2},
\label{p3.26}
\ee
but these expectation values are both BRST invariant and have vanishing
ghost number.

In passing, let us point out a further result of some interest: it is
possible to define new anti-ghost operators $\bg^a$ and $\eta$ by
\be
\beta^a = b^a + \frac{1}{2}\, \eps^{abc}\, \thg c^b c^c, \hs{2}
\eta = \zeta - \frac{1}{3!}\, \eps^{abc}\, c^a c^b c^c.
\label{p3.27}
\ee
These redefinitions preserve the ghost number. Moreover,
one easily establishes the anti-commutation relations
\be
\left[ c^a, \beta^b \right]_+ = \del^{ab},  \hs{2} \left[ \thg, \eta \right]_+ = 1,
\hs{2} \left[ \eta, \bg^a \right]_+= \left[ \thg, c^a \right]_+ = 0,
\label{p3.28}
\ee
with all other anti-commutators vanishing as well. In addition
\be
\beta^a\, \Fg_0 = \eta\, \Fg_0 = 0,
\label{p3.29}
\ee
suggesting that $\Fg_0$ is the actual  Fock vacuum for the
new anti-ghosts $(\eta, \bg^a)$. Unfortunately, it is to be noted that
these antighosts are no longer self-adjoint w.r.t.\ the inner product
(\ref{p3.24}):
\be
\bg^{\dagger}_a = b^a - \frac{1}{2}\, \eps^{abc}\, \thg c^b c^c, \hs{2}
\eta^{\dagger} = \zg + \frac{1}{3!}\, \eps^{abc}\, c^a c^b c^c.
\label{p3.30}
\ee
Hence these operators do not annihilate the conjugate vacuum:
for a general state vector $\Psi$
\be
\lh \Fg_0, \bg^a \Psi \rh = \lh \bg^{\dagger}_a\, \Fg_0, \Psi \rh \neq 0.
\label{p3.31}
\ee
Moreover, the conjugate ghosts have non-trivial anti-commutation
relations with the original anti-ghosts, e.g.:
\be
\left[ \bg^{\dagger}_a, \eta \right]_+ = - \eps^{abc}\, c^b c^c, \hs{2}
\left[ \bg^{\dagger}_a, \bg^b \right]_+ = - \eps^{abc}\, \thg c^c.
\label{p3.32}
\ee
Therefore the ghost variables $(\bg^a,\eta)$ are not of much use in the 
construction of states. Nevertheless, they do provide a good way to 
characterize the ghost dependence of the physical states by the conditions 
(\ref{p3.29}).

\section{Lorenz gauge \label{s.5}}

We will now show, that the problems with the definition of physical states
and inner products sketched in sect.\ \ref{s.4} do not exist in the Lorenz
gauge quantization. The starting point for our analysis is again the classical
theory defined in eqs.\ (\ref{p.1})-(\ref{p.3}), and the representation of 
the nilpotent BRST algebra defined in eq.\ (\ref{p.8}). 
In the $(0 + 1)$-dimensional reduction of the Yang-Mills theory, the Lorenz 
gauge takes the form
\be
\dot{A}_0 = 0.
\label{p5.1}
\ee
A convenient BRST-invariant extension of the classical lagrangean for this 
gauge is
\be
\ba{lll}
L_{Lorenz} & = & \dsp{ L_{YMQM} + N^a \dot{A}_0^a - \frac{1}{2}\, N_a^2
- i \dot{b}^a (D_0 c)^a }\\
 & & \\
 & \simeq & \dsp{ \frac{1}{2}\, (D_0 \bfA^a)^2 + \frac{1}{2}\, (\dot{A}_0^a)^ 2
 - \frac{1}{4}\, (F_{ij}^a)^2 - i \dot{b}^a (D_0 c)^a, }
\ea
\label{p5.2}
\ee
where the last line results from elimination of the auxiliary fields $N^a$.
The corresponding hamiltonian is
\be
\ba{lll}
H_{Lorenz} & = & \dsp{  \frac{1}{2} \lh \bfP^a + g \eps^{abc} A_0^b\, \bfA^c \rh^2
 + \frac{1}{2}\, (P_0^a)^2 + \frac{1}{4}\, (F_{ij}^a)^2
 - \frac{g^2}{2}\, \lh \eps^{abc} A_0^b\, \bfA^c \rh^2}\\
 & & \\
 & & \dsp{  +\, i \lh u^a - i g \eps^{abc}\, A_0^b c^c \rh v^a,}
\ea
\label{p5.3}
\ee
where the canonical momenta are defined by
\be
\ba{ll}
\bfP_a = (D_0 \bfA)^a, & P_0^a = \dot{A}_0^a, \\
 & \\
u^a =- (D_0 c)^a, & v^a = \dot{b}^a.
\ea
\label{p5.4}
\ee
The conserved BRST charge takes the form
\be
\Og = G^a c^a - \frac{ig}{2}\, \eps^{abc}\, c^a c^b v^c + P_0^a u^a,
\label{p5.5}
\ee
with the gauge charges $G^a$ as in eq.\ (\ref{p2.4}). As neither of the
expressions (\ref{p5.3}) and (\ref{p5.5}) suffer from ordering ambiguities,
they can be interpreted directly as quantum operators, with the fundamental
commutation relations given by
\be
\ba{ll}
\left[\bfA_a, \bfP_b \right] = i \del_{ab}\, 1_D, &
\left[ A_0^a, P_0^b \right] = i \del^{ab}, \\
 & \\
\left[ c^a, v^b \right]_+ = \del^{ab}, & \left[ b^a, u^b \right]_+ = \del^{ab}.
\ea
\label{p5.6}
\ee
The quantum equations of motion and the BRST transformations then again
take the form (\ref{p2.3}), (\ref{p2.6}). In the co-ordinate representation, the
BRST-invariant inner product of two wave functions in the full ghost-extended
Hilbert space now takes the form
\be
\lh \Fg, \Psi \rh = i \int \prod_a  db^a dc^a\, \int \prod_a dA_0^a\,
\int \prod_{i,a} dA_i^a\, \Fg^{\dagger}[\bfA, A_0, c, b]\, \Psi[\bfA, A_0, c, b].
\label{p5.7}
\ee
To fix the BRST gauge, we introduce the co-BRST operator
\be
^*\Og = G^a v^a + \frac{ig}{2}\, \eps^{abc} c^a v^b v^c + P_0^a b^a.
\label{p5.8}
\ee
Requiring states to be simultaneously BRST and co-BRST invariant leads
to the conditions
\be
G^a \Psi = 0, \hs{2} \Sg^a \Psi = 0, \hs{2} P_0^a \Psi = 0,
\label{p5.9}
\ee
where
\be
\Sg^a = ig \eps^{abc} c^b v^c,
\label{p5.10}
\ee
is the generator of the rigid $SU(2)$ transformations, which is still an
invariance of the theory, on the conjugate ghosts variables $(c^a, v^a)$.
In contrast to the unitary gauge $A^a_0 = 0$, in the Lorenz gauge the BRST
conditions do not fix the physical states complelety. We can still
impose a further constraint fixing the dependence of physical states on the
anti-ghost variables $(b^a, u^a)$, by requiring states to be rigid $SU(2)$
singlets w.r.t.\ all variables:
\be
\tilde{\Sg}^a \Psi = 0, \hs{2} \tilde{\Sg}^a = i g \eps^{abc}\, b^b u^c.
\label{p5.11}
\ee
Indeed, it is easily checked that $\tilde{\Sg}^a$ is a BRST- and
co-BRST invariant operator; therefore the constraint can be imposed
consistently on all physical states.

The full set of solutions of conditions (\ref{p5.9}) and (\ref{p5.11}) are
wave functions which are $SU(2)$ singlets (i.e., gauge invariant), which
do not depend on $A_0^a$, and whose ghost dependence is constrained
to the form
\be
\ba{lll}
\Psi_{phys}[\bfA,c,b] & = & \dsp{
  \Psi_1[\bfA] + \frac{i}{3!}\, \eps^{abc}\, c^a c^b c^c\, \Psi_2[\bfA]
 + \frac{i}{3!}\, \eps^{abc}\, b^a b^b b^c\, \Psi_3[\bfA] }\\
 & & \\
 & & \dsp{  +\,  \frac{i}{(3!)^2} \lh \eps^{abc}\, c^a c^b c^c \rh
 \lh \eps^{def}\, b^d b^e b^f \rh \Psi_4[\bfA]. }
\ea
\label{p5.12}
\ee
With the standard assignement of the ghost number $+1$ for $c^a$
and $-1$ for $b^a$, requiring the states to have vanishing ghost number
and definite Grassmann parity imposes the further constraint
\be
\Psi_2[\bfA] = \Psi_3[\bfA] = 0.
\label{p5.13}
\ee
Finally, requiring the inner product (\ref{p5.7}) to be positive
definite in the subspace of physical states, we have to fix the space 
of physical states to be represented by factorized wave functions
\be
\ba{lll}
\Psi_{phys} & = & \dsp{
 \frac{1}{\sqrt{2}} \left[ 1 + \frac{i}{(3!)^2} \lh \eps^{abc}\, c^a c^b c^c \rh
 \lh \eps^{def}\, b^d b^e b^f \rh \right] \Psi_M, }\\
 & & \\
 & = & \dsp{  \frac{1}{\sqrt{2}} \lh 1 - i \prod_a (c^a b^a) \rh \Psi_M, }
\ea
\label{p5.14}
\ee
where $\Psi_M$ can be taken as a physical Fock state of the form 
(\ref{p3.13}). Observe that the operator 
$(i/3!)\, \eps^{abc}\, b^a b^b b^c$ plays the same role here as the 
extra ghost $\thg$ in our construction of the states in the unitary
gauge. Obviously, as in the unitary gauge, we can define a ghost 
operator with non-zero vacuum expectation value
\be
\frac{i}{(3!)^2}\, \langle \prod_a (c^a b^a) \rangle = \frac{1}{2},
\label{p5.15}
\ee
but like (\ref{p3.26}) it is BRST invariant and has vanishing ghost number.
Finally, defining the vacuum state of the physical subspace as 
\be
\Fg_0 = \frac{1}{\sqrt{2}} \lh 1 - i \prod_a (c^a b^a) \rh \Psi_0,
\label{p5.16}
\ee
where $\Psi_0$ is the Fock vacuum of the Yang-Mills system, one can
again define ghost operators anihilating $\Fg_0$ by taking
\be
\gam^a = c^a + \frac{i}{2 \cdot 3!}\, \eps^{abc} v^b v^c\, \eps^{def}
 u^d u^d u^f, \hs{2} 
\bg^a = b^a - \frac{i}{2 \cdot 3!}\, \eps^{abc} u^b u^c\, \eps^{def}
 v^d v^d v^f.
\label{p5.17}
\ee
As might be expected from our previous analysis, these operators
are not self-adjoint and do not define a good basis for a complete
Fock-space construction in the ghost sector. Nevertheless, the conditions
\be
\gam^a \Fg_0 = \bg^a \Fg_0 = 0
\label{p5.18}
\ee
provide a convenient way to characterize the physical ghost vacuum.

Finally we should remark, that in the physical subspace the integration
over $A_0^a$ is of course divergent in the absence of damping, as the
physical wave functions are $A_0$-independent. This divergence can
be absorbed in a wave-function renormalization factor
\be
N = \frac{1}{\sqrt{\int \prod_a dA_0^a}}.
\label{p5.19}
\ee
Knowing this, we can remove the $A_0^a$ from the physical inner product
and effectively set $N = 1$; we observe, that $N$ is BRST-invariant,
and the procedure does not jeopardize the BRST-invariance of the
integration measure.

\section{Discussion \label{s.6}}

In this paper we have shown, that although the physical content
of the $(0+1)$-dimensional Yang-Mills theory is clearest in the
unitary gauge $A_0^a = 0$, the BRST quantization works in a more
straightforward way in the Lorenz gauge $\dot{A}_0^a = 0$. An
important part of the discussion and analysis was based on the
construction of a BRST-invariant inner product w.r.t.\ which the
BRST charge $\Og$ is self-adjoint.

To get a little more algebraic and geometric insight into the
constructions, consider again the unitary gauge, in which a
general state is represented by a wave function
\be
\Psi[c] = \psi + c^a \psi_a + \frac{i}{2!}\, c^a c^b \psi_{ab}
 + \frac{i}{3!}\, c^a c^b c^c \psi_{abc}.
\label{p6.1}
\ee
Defining the dual wave function
\be
\tilde{\Psi}[c] = \tilde{\psi} + c^a \tilde{\psi}_a + \frac{i}{2!}\, c^a c^b
\tilde{\psi}_{ab} + \frac{i}{3!}\, c^a c^b c^c \tilde{\psi}_{abc}.
\label{p6.1.1}
\ee
with components
\be
\ba{ll}
\dsp{ \tilde{\psi} = \frac{1}{3!}\, \eps_{abc} \psi_{abc}, }&
\dsp{ \tilde{\psi}_a = \frac{1}{2!}\, \eps_{abc} \psi_{bc}, }\\
 & \\
\dsp{ \tilde{\psi}_{ab} = \eps_{abc} \psi_c, }& \dsp{
\tilde{\psi}_{abc} = \eps_{abc} \psi, }
\ea
\label{p6.2}
\ee
we recognize that the physical states (\ref{p3.19}) are characterized
as the self-dual states $\tilde{\Psi} = \Psi$, such that the inner product
(\ref{p2.8}) becomes
\be
\ba{lll}
i \int dc^1 dc^2 dc^3\, \Psi^{\dagger} \Psi  & = & \dsp{
 \frac{1}{3!}\, \eps_{abc} \lh \psi_{abc}^{\dagger} \psi + \psi^{\dagger} \psi_{abc}
 + 3 \psi^{\dagger}_a \psi_{bc} + 3 \psi^{\dagger}_{ab} \psi_c \rh }\\
 & & \\
 & = & \dsp{ 2 \psi^{\dagger} \psi + 2 \psi^{\dagger}_a \psi_a. }
\ea
\label{p6.3}
\ee
In particular,  with $\tilde{\psi} = \psi = \sqrt{2}\, \psi_M$ and
$\psi_a = \tilde{\psi}_a = 0$, this reduces to
\be
i \int dc^1 dc^2 dc^3\, \Psi^{\dagger}[M] \Psi[M] = \psi^{\dagger}_M\, \psi_M.
\label{p6.4}
\ee
Hence this inner product is positive definite for physical states.
Of course, one can also consider the anti-self dual states $\tilde{\Psi}
 = -\Psi$, which then have a negative definite norm. This should not
surprise us, as the existence of a self-adjoint nilpotent  BRST operator
$\Og^2 = 0$ is possible only in a space with indefinite norm. The important
point is, that the space of physical states should have positive norm,
and that is realized in the subspace of self-dual states.

Generalization of this discussion to the Lorenz gauge is simple. Each
component in the wave-function expansion (\ref{p6.1}) now is a function
of the additional ghost variables $b_a$, and we can again distinguish
between components which are self-dual or anti-self-dual w.r.t.\ the
expansion in $b_a$. In this formulation the physical states are then
identified with the wave functions for which the components of zero
ghost-number are completely self dual, i.e.\ self-dual both with respect
to the $c$-ghost duality and with respect to the $b$-ghost duality.

We have discussed in particular the case of $SU(2)$ Yang-Mills theory.
The generalization to $SU(N)$ is straightforward; with $r = N^2 - 1$
generators, and the same number of ghost and anti-ghost variables,
the self-dual physical states in the unitary gauge are of the form
\be
\Psi[c] = \frac{1}{\sqrt{2}} \lh 1 + \frac{i^{[r/2]}}{r!}\,
 \eps^{a_1\, ...\, a_r}\, c^{a_1} ... c^{a_r} \rh \psi_M.
\label{p6.5}
\ee
For odd $r$ (even $N$), both ghost number and Grassmann parity of
the wave functions are ill-defined; for even $r$ (odd $N$), it is only
the ghost number which is violated. In both case, introduction of
a singlet ghost $\thg$ with ghost number $n_g(\thg) = - r$ solves
the problems. On the other hand, in the Lorenz gauge this is
taken care of automatically by the anti-ghost variables, as the
operator
\be
\frac{i^{[r/2]}}{r!}\, \eps^{a_1\, ...\, a_r}\, b^{a_1} ... b^{a_r}
\label{p6.6}
\ee
has the same quantum numbers and plays the same role.

Finally we note, that as we have constructed precisely one BRST-invariant
wave function for each physical state, in the supersymmetric extension
the computation of the Witten index \ct{yi}-\ct{vanbaal} is not affected by
including the ghost degrees of freedom in the appropriate way. \\

{\bf{Acknowledgements.\/}} We would like to thank J. Wosiek. This work is part of the programme FP52 of the Foundation for 
Research of Matter (FOM). The work of A.F. was also supported by a Basque Government 
grant until $01$/$10$/$04$.

\np

\end{document}